\documentclass[12pt]{article}
\usepackage{amsmath}
\newcommand{\bibspace}{\vspace{1.0ex}\noindent}
\begin{document}
\begin{center}
\LARGE
\textbf{Multiplicity in Everett's
interpretation of quantum mechanics}\\[1cm]
\large
\textbf{Louis Marchildon}\\[0.5cm]
\normalsize
D\'{e}partement de chimie, biochimie
et physique,\\
Universit\'{e} du Qu\'{e}bec,
Trois-Rivi\`{e}res, Qc.\ Canada G9A~5H7\\
email: louis.marchildon$\hspace{0.3em}a\hspace{-0.8em}
\bigcirc$uqtr.ca\\
\end{center}
\medskip
%
%\begin{frontmatter}
%
%\title{Multiplicity in Everett's
%interpretation of quantum mechanics}
%
\author{Louis Marchildon}
%
%\address{D\'{e}partement de chimie, biochimie
%et physique, Universit\'{e} du Qu\'{e}bec,
%Trois-Rivi\`{e}res, Qc.\ Canada G9A~5H7
%\ead{louis.marchildon@uqtr.ca}}
%
\begin{abstract}
Everett's interpretation of quantum mechanics
was proposed to avoid problems inherent in
the prevailing interpretational frame.  It
assumes that quantum mechanics can be applied
to any system and that the state vector always
evolves unitarily.  It then claims that whenever
an observable is measured, all possible results
of the measurement exist.  This notion of
multiplicity has been understood in different
ways by proponents of Everett's theory.  In
fact the spectrum of opinions on various
ontological questions raised by Everett's
approach is rather large, as we attempt to
document in this critical review.  We
conclude that much remains to be done to
clarify and specify Everett's approach.
\end{abstract}
\section{Introduction\label{intro}}
Everett's `relative state formulation of
quantum mechanics' (1957a), also known as `the
many-worlds interpretation,' was proposed almost
sixty years ago.  It remained little more than
a curiosity for a couple of decades, but has
from then on attracted sustained attention.
Its current status was synthesized and
criticized in the remarkable recent monograph
by Saunders \emph{et al.} (2010),  while Byrne
(2010) uncovered at the same time the
fascinating story of its genesis.

Everett's original motivation was to provide
a formulation of quantum mechanics that would
avoid problems inherent in the then current
interpretational frame, which drew both from
the Copenhagen distinction between the quantum
and the classical and from the Dirac--von Neumann
collapse of the state vector.  The main
problem comes in so-called measurement
situations which, in view of later discussions,
we presently formulate on a specific example.

Let $|a_1\rangle$ and $|a_2\rangle$ be an
orthonormal basis of a two-dimensional
Hilbert space, associated for instance with
a spin-$\frac{1}{2}$ particle. Let an
apparatus be designed so as to measure an
observable $\Sigma_z$ given by
\begin{equation}
\Sigma_z = |a_1\rangle \langle a_1|
- |a_2\rangle \langle a_2| .
\end{equation}
Matrix elements of $\Sigma _z$ in the above
basis are given by the Pauli matrix
denoted by~$\sigma_z$.

Let $|\alpha_0\rangle$ denote the normalized
initial state of the apparatus.  In a
nondestructive measurement, the interaction
between the particle and apparatus is assumed
to be effected by a unitary operator
such that
\begin{equation}
|a_1\rangle |\alpha_0\rangle \rightarrow
|a_1\rangle |\alpha_1\rangle , \qquad
|a_2\rangle |\alpha_0\rangle \rightarrow
|a_2\rangle |\alpha_2\rangle ,
\label{invar}
\end{equation}
where the arrow symbolizes time evolution and
$|\alpha_1\rangle$ and $|\alpha_2\rangle$
are orthogonal pointer states.  Let $c_1$
and $c_2$ be complex coefficients (satisfying
$|c_1|^2 + |c_2|^2 = 1$ for normalization).
Because the unitary evolution is linear,
we must have
\begin{equation}
(c_1 |a_1\rangle + c_2 |a_2\rangle)
|\alpha_0\rangle \rightarrow
c_1 |a_1\rangle |\alpha_1\rangle
+ c_2 |a_2\rangle |\alpha_2\rangle .
\label{semifinal}
\end{equation}
Thus the interaction has transformed an initial
product state into a final entangled state.

If we look carefully at the right-hand side
of~(\ref{semifinal}), we see that the
apparatus finds itself in a superposition
of different pointer states.  This is
highly counterintuitive.  It doesn't help
to introduce an observer $O$ to read the
pointer and evolve from a normalized
state $|O_0\rangle$ to a state $|O_1\rangle$
if he reads $\alpha_1$, or to a state
$|O_2\rangle$ if he reads $\alpha_2$.  For
if $O$ is treated quantum-mechanically, the
combined particle-apparatus-observer system
evolves as
\begin{align}
(c_1 |a_1\rangle + c_2 |a_2\rangle)
|\alpha_0\rangle |O_0\rangle &\rightarrow
(c_1 |a_1\rangle |\alpha_1\rangle
+ c_2 |a_2\rangle |\alpha_2\rangle) |O_0\rangle
\notag\\
&\rightarrow
c_1 |a_1\rangle |\alpha_1\rangle |O_1\rangle
+ c_2 |a_2\rangle |\alpha_2\rangle |O_2\rangle .
\label{final}
\end{align}
At the end of the process, the observer
therefore finds himself in a superposition.

The Copenhagen solution of the measurement
problem consists essentially in pointing out
that an apparatus is necessarily classical,
and therefore cannot be a quantum system as
hypothesized before~(\ref{invar}).  The upshot
is that the apparatus registers either $\alpha_1$
or $\alpha_2$, but not both.  Copenhagen
adherents bear the burden of precisely
specifying where the quantum-classical
boundary lies.\footnote{Or, they have to
live with the fact that an apparatus
behaves classically if it performs a
measurement, and quantum-mechanically if
it is itself measured by a super apparatus.}

The Dirac--von Neumann solution of the
measurement problem consists in introducing
a break in the unitary evolution of the
state vector.  It doesn't matter whether
the break occurs after the system-apparatus
coupling or after the system-apparatus-observer
coupling.  In the former case for instance,
the break occurs as
\begin{align}
& c_1 |a_1\rangle |\alpha_1\rangle
+ c_2 |a_2\rangle |\alpha_2\rangle \notag\\
& \qquad \rightarrow
|c_1|^2 (|a_1\rangle \langle a_1|)
(|\alpha_1\rangle \langle \alpha_1|)
+ |c_2|^2 (|a_2\rangle \langle a_2|)
(|\alpha_2\rangle \langle \alpha_2|) .
\label{collapse}
\end{align}
The right-hand side of~(\ref{collapse})
represents a proper mixture where the
pointer shows $\alpha_1$ with probability
$|c_1|^2$ and $\alpha_2$ with probability
$|c_2|^2$.  Von Neumann (1955) called the
break Process~1 (the collapse of the
state vector), contrasting it with Process~2,
the unitary Schr\"{o}dinger evolution.
Proponents of collapse bear the burden of
precisely specifying how and in what
circumstances Process~1 occurs.

Everett's proposal was to
take~(\ref{semifinal}) and~(\ref{final})
at face value.  This means that, in some
sense, all outcomes of a measurement have
the same ontological status.  The resulting
interpretation of quantum mechanics would
not require the presence of a classical
apparatus or observer external to
a quantum system.  Quantum mechanics could
therefore, in principle, be applied to the
whole universe.  Moreover, the theory would not
require an explicit collapse of the state vector.
The state vector would only evolve through
Process~2, that is, through Schr\"{o}dinger's
unitary evolution.

I have argued elsewhere (Marchildon, 2011)
that the most important problem related
to Everett's work is the understanding of
multiplicity.\footnote{This assessment is
not consensual.  See for instance the
extensive discussion of probability in
Saunders \emph{et al.} (2010).  Schwindt
(2012) argues that the preferred-basis
problem has a component which is not solved
by decoherence, while Baker (2007)
correspondingly emphasizes how probability,
decoherence and the preferred basis are
interconnected.}
It is clear that for Everett, all possible
results of a quantum experiment exist.
In what sense they do exist, however,
was not made precise in Everett's published
work.  Perhaps the closest he comes to
specifying this is found in the following
quote (Everett, 1957a, p.~459):\footnote{Unless
specified otherwise, emphasis is always in
the original.}
\begin{quote}
We thus arrive at the following picture:
Throughout all of a sequence of observation
processes there is only one physical system
representing the observer, yet there is no
single unique \emph{state} of the observer
(which follows from the representations of
interacting systems). Nevertheless, there is
a representation in terms of a
\emph{superposition}, each element of which
contains a definite observer state and a
corresponding system state. Thus with each
succeeding observation (or interaction), the
observer state `branches' into a number of
different states.  Each branch represents a
different outcome of the measurement and the
\emph{corresponding} eigenstate for the
object-system state. All branches exist
simultaneously in the superposition after
any given sequence of observations.
\end{quote}

Its title notwithstanding, this paper
does not aim at carefully reconstructing
the thread of Hugh Everett III's views
on multiplicity from all available
published and unpublished sources,
something on which substantial progress
has been achieved in recent
years.\footnote{See for instance Osnaghi
\emph{et al.} (2009), Byrne (2010),
Barrett (2011), Barrett and Byrne (2012)
and Barrett (2014).} The purpose
of the paper is to provide a fairly
exhaustive review of the ways people have
tried in subsequent years to understand
multiplicity in Everett's
framework.\footnote{Inevitably, there is some
overlap between this paper and Barrett (1999).
But Barrett's excellent review of Everett's
approach is now more than 15 years old.
Earlier syntheses can be found in Whitaker
(1985) and Kent (1990).  The former is concerned
with the relation between Everett's approach and
the EPR problem.  The latter is an influential
critique of many-worlds interpretations, aimed
mainly at the probability and preferred-basis
problems.}  They can be
broadly divided into three groups: many worlds,
many minds, and decoherent sectors of the wave
function.  We shall see that these approaches
further subdivide, and that they all raise
important questions that have not been
completely answered.  We shall not presume to
make any final assessment on Everett's program,
but we shall conclude at least that there is
still much to be clarified and specified in it.
%
%\newpage
\section{Many worlds\label{mw}}
The phrase `many worlds' evokes the genuine
existence of a great number of more or less
similar copies of the world we live in.
The idea that the world literally splits
into a number of (initially) slightly different
real copies of itself was popularized by DeWitt.
Indeed (DeWitt, 1970, p.~33):
\begin{quote}
This universe is constantly splitting into
a stupendous number of branches, all resulting
from the measurementlike interactions between
its myriads of components. Moreover, every
quantum transition taking place on every star,
in every galaxy, in every remote corner of
the universe is splitting our local world
on earth into myriads of copies of itself
[...] I still recall vividly the shock I
experienced on first encountering this
multiworld concept.  The idea of $10^{100+}$
slightly imperfect copies of oneself all
constantly splitting into further copies,
which ultimately become unrecognizable,
is not easy to reconcile with common sense.
\end{quote}

According to this quote, the split occurs at
the atomic, that is, at the microscopic
level.  Every time an interaction produces
entanglement, whether or not macroscopic
objects are involved, a split presumably
occurs.  This goes beyond what we find
in Everett, at least in print.  There are
good indications (Byrne, 2010; Bell, 1976,
p.~95; Deutsch, 1996, p.~223) that Everett,
just like DeWitt, had in mind a genuine
split.\footnote{But Osnaghi \emph{et al.}
(2009, p.~107) note that throughout
Everett's writings, the terms `real,'
`reality,' `actual,' `branching process'
and `branches' appear systematically in
quotes.  In an analysis of Everett's
attitude towards `worlds,' Barrett (2011,
p.~286) points out that ``Everett
believed that his theory neither
required nor supported any special
metaphysical commitments.''
See also Zeh (2014).}
Unlike DeWitt, however, he
applied it only in contexts where there
is an interaction involving something like
a macroscopic apparatus.  Moreover, he
proposed rather specific conditions under
which measurements occur (Everett, 1957b,
pp.~10, 53, 55):
\begin{quote}
[A]ll measurement and observation processes
are to be regarded simply as interactions
between observer and object-system which
produce strong correlations.

[A] measurement is simply a special case of
interaction between physical systems---an
interaction which has the property of
\emph{correlating} a quantity in one subsystem
with a quantity in another.

We shall therefore accept the following
definition.  An interaction H is a
measurement of A in $\mbox{S}_1$ by B in
$\mbox{S}_2$ if H does not destroy the marginal
information of A (equivalently: if H
does not disturb the eigenstates of A in
the above sense) and if furthermore the
correlation \{A,B\} increases toward
its maximum \mbox{[\ldots]} with time.
\end{quote}
In other words, a measurement transforms
a product state of a quantum system and
an apparatus into a maximally-entangled
state, in a nondestructive way (i.e.\
system eigenstates are left invariant).

Already at this stage, we are confronted
with two different views of splitting,
which give rise to two questions:
\begin{enumerate}
\item Does the split occur with all
microscopic entanglement-producing
interaction, or only in macroscopic
measurementlike contexts?
\item Does the split occur in every
possible basis in which the entangled
state can be expressed, or just in one
basis?
\end{enumerate}

With respect to the second question,
it seems that it would be impossible
to recover appearances if a genuine
split occurred along any basis.  This
would mean real worlds with macroscopic
apparatus, and even human observers,
in quantum superpositions.  We will see,
however, that such a conclusion may not
hold in other interpretations of
Everett's theory.  Butterfield (1995, p.~133)
pointed out that branching along arbitrary
bases may be coherent, but it presumably
implies sacrificing natural meshing
conditions between the branches of different
quantities, in order to avoid no hidden
variable theorems.

So it seems that with a genuine split,
there must be some preferred basis.
Ballantine (1973) pointed out that
if the split occurs at the atomic level,
there is no natural definition of a
branching representation.  Note that it
doesn't help to propose that branching
occurs along states that are left invariant
by the interaction, like
$|a_1\rangle$ and $|a_2\rangle$ in~(\ref{invar}).
For this nondestructive interaction is
very special, and one can easily
contemplate entanglement-producing
interactions that leave no states
invariant.  If the split only occurs
at a macroscopic level, decoherence no
doubt helps in selecting a preferred basis.
But following Bell, Everett can be
criticized for not
making the concept of instrument reading
precise (Bell, 1976, pp.~96--97):
\begin{quote}
[I]f instrument readings are to be given such a
fundamental role should we not be told more
exactly what an instrument reading is
\mbox{[\ldots]} [F]undamental physical theory
should be so formulated that such artificial
divisions are manifestly inessential.
\end{quote}
We see that if the split occurs only with
macroscopic apparatus, the problem of
specifying what an apparatus is looks very
much like the problem of clarifying the
classical-quantum distinction in the
Copenhagen interpretation.

Vaidman (1998, p.~251) envisions a microscopic
split in connection with neutron
interferometry: ``My proposal is that during
the period of time the neutron wave function
is inside the interferometer \emph{there are
two neutron worlds}[.]''  But we will see later
that he proposes rather robust conditions
for a macroscopic split.  Bell (1986, p.~193)
had summarized the ambiguity between
microscopic and macroscopic split in the
following way: ``One is given no idea of how
far down towards the atomic scale the
splitting of the world into branch worlds
penetrates.''

There are several distinct ways to view
splitting, even if it is restricted
to occur in connection with macroscopic
events only.

Healey (1984) considers that splitting
occurs upon a measurementlike interaction.
According to him, the simplest way to view
splitting (say, into $n$ copies) is that
every elementary system splits into $n$
copies, in usual ordinary space.  But this,
for Healey, is hardly defensible, for
mass-energy would also be multiplied
and we would presumably be aware of the
overcrowding of space.  Accordingly, the
split he introduces can become acceptable
in two different ways:
\begin{enumerate}
\item The physical systems do not split,
only their states do.
\item Not only systems, but space itself
splits.  The resulting systems may be viewed
as living in a higher-dimensional space.
\end{enumerate}
Healey develops the first way into lines that
anticipate what we shall elaborate in
Sect.~\ref{deco}. With respect to the second
way, he claims (pp.~598--599) that
\begin{quote}
within the framework of ordinary
non-relativistic quantum mechanics,
a version of the many-worlds interpretation
according to which space and its constituent
quantum systems split on quantum measurement
offers no interpretative advantages over a
version in which the actual outcome of a
measurement is only one of many
\emph{possible} states of the world in
space.
\end{quote}
Healy then analyses the actual one-world
version he has in mind.  It turns out to be
related to modal interpretations (Vermaas,
1999) and to share some of their problems.
Note that in the context of genuine
splitting, the relevance of retaining many
worlds has often been questioned.  Indeed
(Kent, 2012, p.~423):
\begin{quote}
As Bell \mbox{[\ldots]} and (probably many)
others \mbox{[\ldots]} have noted, if the
branching of worlds were precisely and
objectively defined, a many worlds
interpretation would seem unnecessarily
extravagant. Given a precisely defined
branching structure, we can just as easily
define a \emph{one world} interpretation
of quantum theory.
\end{quote}

Several investigators have proposed
reasonably specific definitions of
what should be counted as a world.
Vaidman (2014) clearly commits to
a definition along macroscopic lines.
He proposes that the quantum state of
a world is given by
\begin{equation}
|\Psi_{\text{world}}\rangle
= |\Psi\rangle_{\text{object}\,1}
|\Psi\rangle_{\text{object}\,2} \ldots
|\Psi\rangle_{\text{object}\,N} |\Phi\rangle,
\label{world}
\end{equation}
where each `object' is macroscopic and
$|\Phi\rangle$ represents the quantum state
of all the particles that do not constitute
objects.  In the 2002 version of the same
Encyclopedia article, he also considered the
possibility that only the object states
directly perceived by sentient beings appear
in~(\ref{world}), with $|\Phi\rangle$
representing everything else. 

Butterfield (1995) had a similar representation,
where the $|\Psi\rangle_{\text{object}\,i}$
are decohering states of macroscopic objects
and $|\Phi\rangle$ is the overall relative
state.  Discussing how splitting and
multiplicity (or `plurality') can apply
to branches, he concludes that
decoherence removes the need of physical
splitting.  In Butterfield (2001), formally
similar definitions of worlds are used both
for many-worlds and decoherence approaches.

Graham (1973) interprets Everett's as a
world-splitting theory, where there is one
world corresponding to each possible result
of a measurement.  In an attempt to
recover the Born rule from Everett's
approach, he applies the split to an
observer reading a macroscopic apparatus
intended to measure the relative frequency
operator on a collection of identically
prepared systems.  It is not clear if the
split also applies to a single measurement
of a microscopic observable.

Here again, precision in definitions may
reduce the appeal of a many-worlds
approach.  Indeed (Barrett, 1999, p.~157):
\begin{quote}
DeWitt and Graham's world-splitting rule
tells us that worlds split whenever a
measurement-like interaction occurs, but
they never explain precisely what counts as
a measurement-like interaction; rather, this
is determined by one's choice of a preferred
basis in the theory, which is never made
explicit.  But note that if we \emph{did know}
what it took to count as a measurement-like
interaction here, then one would be able
to solve the measurement problem in the
standard collapse theory by stipulating
that the global wave function collapses
whenever precisely that sort of interaction
occurs.
\end{quote}

The next important question that comes up
with genuine splitting is whether the
split is irreversible or, equivalently,
whether each branch henceforth evolves
completely independently from the others.
In a collective reply to DeWitt (1970),
Gerver points out (Ballantine
\emph{et al.}, 1971, p.~40) that
\begin{quote}
if it is possible for the universe to split
into two slightly different realities by a
quantum-mechanical event, then surely it is
equally possible for two slightly different
universes to become identical in the same
manner.
\end{quote}
DeWitt (1970, p.~35) also believes that in
principle the split can be undone, ``by
bringing the apparatus packets back together
again.''  Although Everett (1957b, p.~97)
speaks of ``an essential irreversibility
to Process~1,'' that irreversibility is
only apparent within a theory ``which
recognizes only Process~2'' (p.~98).
As Barrett (2014) points out, branches are
real for Everett because they are always
in principle detectable (Everett 1957b,
p.~107):
\begin{quote}
It is therefore improper to attribute
any less validity or `reality' to any
element of a superposition than any
other element, due to [the] ever present
possibility of obtaining interference
effects between the elements.  All elements
of a superposition must be regarded as
simultaneously existing. 
\end{quote}

It should be pointed out that undoing the
split would involve a rather peculiar
process, even if we admit that splitting
occurs in purely microscopic interactions.
Suppose that a split has occurred after
interaction~(\ref{semifinal}), where
$|\alpha_1 \rangle$ and $|\alpha_2 \rangle$
represent microscopic states.  For a time
both terms on the right-hand side
of~(\ref{semifinal}) are different worlds.
Now it may happen that $|\alpha_1 \rangle$
evolves into $|\alpha_0 \rangle$ in world~1,
while $|\alpha_2 \rangle$ evolves into
$|\alpha_0 \rangle$ in world~2, with
$|a_1 \rangle$ and $|a_2 \rangle$ unchanged.
Then the global state would revert to
the left-hand side of~(\ref{semifinal}).
At that moment, world~1 and world~2
would unsplit, or recombine into just
one world.  Zeh (1970) points out that
someone in a given branch cannot estimate
the probability of inverse branching.

Several authors have pointed out that
splitting into completely independent
worlds is incompatible with
Schr\"{o}dinger dynamics:
\begin{quote}
Apparent collapse has also been described
as `world-splitting,' and, as several authors
have warned \mbox{[\ldots]} it is tempting to take
world-splitting as a physical process and
thus effectively return to the Copenhagen
interpretation. (Donald, 1995, p.~532)

If one supposes that the state of a world
determines the behaviour of physical systems
in that world in so far as their behaviour
is determined, then the splitting-worlds
theory is incompatible with the usual
linear dynamics[.] (Barrett, 1999, p.~160)
\end{quote}
A similar conclusion can be drawn from
Albert and Barrett (1995), who consider the
measurement of an observable pertaining to
an apparatus that has already made a spin
measurement.\footnote{Different objections to
Everett branching can be found in Gauthier
(1983) and Jekni\'{c}-Dugi\'{c} \emph{et al.}
(2014).}

The difficult problem of recombination can,
however, be avoided.  Deutsch (1985) suggests
that there are infinitely many worlds
(he calls them `universes') at
any time.  Their number neither increases nor
decreases.  In measurement contexts the
set of all worlds is partitioned in as
many branches as there are possible
measurement results, the measure of worlds
in each branch corresponding to the
probability of the associated observable
value.  Thus we have bifurcation instead
of splitting.\footnote{A similar view has
been developed within the approach to
Everett that we investigate in
Sect.~\ref{deco}.  See Saunders (2010)
and Wilson (2011).  Branches that coincide
in the past are said to `overlap,' whereas
they `diverge' if in the past they are
only qualitatively identical.}
To the measured observable
Deutsch attempts to associate an
interpretation basis in the Hilbert space,
within the quantum formalism.  Deutsch
claims that the probabilistic interpretation
is now truly built in, although Butterfield
(2001) argues that it is difficult to make
sense of probability in Deutsch's approach.

Note that bifurcation avoids a problem
that several people see with splitting,
namely, a dramatic increase in
mass-energy, in violation with the known
conservation laws.  No such thing can happen
if the number of worlds does not change.

A continuous infinity of worlds is also
postulated by Bostr\"{o}m (2012), who
tries to combine the approaches of
Everett and Bohm.  Bostr\"{o}m considers
the configuration space of $N$ pointlike
particles and associates, at a given time,
a distinct world with each configuration.
However, he identifies worlds which differ
only by the permutation of two identical
particles.  Note that many worlds along
Bohmian lines were also considered by
Bell (1976, 1981), Tipler (2006) and Valentini
(2010).  Bostr\"{o}m attempts to answer
the criticism that multiplicity in this
case is entirely artificial.  He further
develops his ideas in Bostr\"{o}m (2015).

In perspicuous early analyses of Everett's
approach, Bell suggested that the real novel
element in Everett's theory is the repudiation
of the concept of past: 
\begin{quote}
Everett [...] tries to associate each
particular branch at the present time with
some particular branch at any past time in
a tree-like structure, in such a way that
each representative of an observer has
actually lived through the particular past
that he remembers.  [This] attempt does not
succeed \mbox{[\ldots]}
and is in any case against the spirit
of Everett's emphasis on memory contents as
the important thing.  We have no access to
the past, but only to present memories.
(1976, p.~95)

Keeping the instantaneous configurations,
but discarding the trajectory, is the
essential (in my opinion) of the theory
of Everett.  (1981, p.~133)
\end{quote}
Butterfield (1995) also argues
that the absence of history tends to
support plurality.

If splitting into many worlds is a
physical process, several specific
questions arise.  Thus Lockwood
(1989, p.~226) asks ``at what point in
the von Neumann chain does the split or
decomposition occur, and on what spacelike
surface?''  Tipler (1986, p.~206) had
partly answered the second query:
\begin{quote}
[M]any presentations of the MWI [many-worlds
interpretation] have made it appear more
counter-intuitive than it really is.
For example, many accounts assert that ``the
entire universe is split by a measurement.''
This is not true.  Only the observed/observer
system splits; only that restricted portion
of the universe acted on [by] the measurement
operator $M$ splits.
\end{quote}
It is not clear where the two physical
instances of the apparatus sit if the rest
of the universe is not split.

Squires (1988, p.~16) raises the question
``what can `splitting' possibly mean in this
context---what moves away from what?---and
in what `space'?'' To such interrogations,
Tappenden (2000, p.~113) replies:
\begin{quote}
This all suggests that what is involved
in unpacking Everett's proposal is the
addition of a further dimensionality
to standard four-dimensional spacetime[.]
\end{quote}
But Vaidman (2014) claims that
the worlds of the many-worlds interpretation
``exist in parallel at the same space and
time as our own.'' We will come back to this in
Sect.~\ref{deco}.

In an early discussion of Everett's and
DeWitt's ideas, Smolin (1984) first
introduced what he calls the ``minimal
relative state interpretation'' (MRS).
In the MRS, the right-hand side
of~(\ref{semifinal}) is not interpreted
as the actual state of the composite
system, but as the list of contingent
statements like `If the apparatus reads
$\alpha_1$, then the quantum system's
observable $\Sigma_z$ has value~$+1$.'
Smolin then shows that the MRS has several
of the advantages that the full-fledged
MWI has.  In the MWI, the right-hand side
of~(\ref{semifinal}) represents the actual,
objective state of affairs of the composite
system, where all outcomes are actual
(p.~445):
\begin{quote}
[I]f we wish to regard our having observed
a particular $a_i$ to be the outcome of the
experiment as an actual event in the world then
it follows that we must also regard our having
observed each of the other possibilities as
also being actual events in the world.
\end{quote}
But Smolin also expresses caution about
splitting (pp.~447--448):
\begin{quote}
One usually introduces at this point in the
discussion the expression that the different
statements are all true, but each is true of
a different `branch' of the wavefunction
\mbox{[\ldots]} The disadvantage of using the
language of branches is that it suggests that
some kind of dynamical mechanism is taking
place, in addition to the evolution of the
wavefunction, in which, as time goes on and
initially isolated systems come together and
interact, the universe is `splitting' into
more and more `branches.'
\end{quote}
This makes it more difficult to understand
what Smolin has in mind with the actuality
of all outcomes, unless it is interpreted
as in the next sections.
%
%\newpage
\section{Many minds\label{mm}}
The first full-fledged formulation of the
idea that the split involves the mind
rather than the world seems to be the one of
Albert and Loewer (1988).  Yet some passages
in Everett's published work can be construed
in that way. For instance we read
(Everett, 1957b, p.~10) that
\begin{quote}
after the interaction has taken place there
will not, generally, exist a single observer
state.  There will, however, be a superposition
of the composite system states, each element
of which contains a definite observer state
and a definite relative object-system state
[...] Thus, each element of the resulting
superposition describes an observer who
perceived a definite and generally different
result, and to whom it appears that the
object-system state has been transformed into the
corresponding eigenstate.  In this sense the
usual assertions of Process 1 appear to hold
on a subjective level to each observer
described by an element of the superposition.
\end{quote}

Cooper and Van Vechten (1969, pp.~1217--1218),
who point out the similarity of their views
to Everett's, come rather close to the idea
of many minds:
\begin{quote}
[O]ur knowledge that our own mind is in some state
need not be reflected in the wave function.
Rather, it is expressed in the way we pose the
question.  That a system is in the state $U$
is equivalent to the statement that all coupled
`good' systems and sane minds will agree that
the system is in the state $U$ \mbox{[\ldots]}
The wave function may contain a superposition of
$U$ and $L$ but there is no manifestation of this
to anyone---including the mind described by the
wave function---unless interference can occur.
\end{quote}

Zeh (1970, 1981) anticipates the many-minds
view, and attributes the idea to Everett.  Healey
(1984) also suggests that Everett can be
understood along that way, while the general
idea is further anticipated in Albert (1986)
and Squires (1987, 1988, 1991).

Albert and Loewer motivate their introduction
of many minds by listing three problems with
the splitting-worlds view: (i) the fact that,
as they see it, it entails the nonconservation
of mass-energy; (ii) the difficulty to
understand probability within a deterministic
theory; and (iii) the choice of a preferred
basis.  They then point out
that if the Schr\"{o}dinger equation is
always valid, the brain states of an observer,
in a measurement context, can evolve into
a superposition.  Belief states, however,
are never superposed, for an observer never
believes he is in a superposition of seeing
$|\alpha_1\rangle$ and seeing $|\alpha_2\rangle$.
Therefore, belief states cannot be wholly
physical.  Albert and Loewer thus commit
themselves to some form of dualism.

Albert and Loewer then introduce minds in
two different ways, the latter being their
favorite.  In the single-mind view (SMV),
each observer has one mind.  The state
vector after measurement is given
by~(\ref{final}), but the observer's mind
is associated either with
$O_1$ or (exclusive) with $O_2$,
with probabilities $|c_1|^2$ and $|c_2|^2$
respectively.  Albert and Loewer point out
that the nonphysicalism in this case is
rather acute, for mind states do not even
supervene on physical states.  Moreover,
the SMV gives rise to what has become
known as the mindless-hulk problem (Albert,
1992).  This consists in the fact that
after the split, all brain states but one
are mindless.  Thus in EPR contexts, Alice's
single mind and Bob's single mind can
end up in different branches, unless there
are strong nonlocal correlations between
minds.

In the many-minds view (MMV), every observer has
associated with it an infinite set of minds.
In~(\ref{final}), a fraction $|c_1|^2$ of
minds become associated with $O_1$ and
a fraction $|c_2|^2$ with $O_2$.
Minds are associated with brain states
but are not subject to superposition.
In spite of this nonphysicalism, mental
states are supervenient on brain states.
Although the time evolution of each mind
is probabilistic, the time evolution of
the set of all minds is deterministic.
Albert and Loewer claim that the many-minds
view is local.  Although they do not fully
commit to it, they favor the transtemporal
identity of individual minds.  Since the
transtemporal identity of individual
minds has no ground in physical facts,
this makes up for a definitely dualistic
view.  Saunders (1996b) notes
that there is no mindless hulk problem if
the transtemporal identity of minds is
rejected.

Barrett (1995) rehabilitates the SMV,
provided the mental dynamics is
suitably constrained by the linear
physical dynamics.

Squires (1988) also proposed what amounts
to the SMV.  Later Squires (1991, p.~285)
introduced a concept of universal
consciousness, to avoid the mindless hulk
problem (although he didn't use the term):
\begin{quote}
The obvious way of satisfying this requirement
[that the choices made by Jack's and Jill's
conscious minds be correlated] is to assume
some sort of universality of consciousness
so that when, for example, Jack's conscious
mind is aware of the result, then
`Conscious Mind' is also aware of it
\mbox{[...]} The [alternative] idea \mbox{[...]}
is that, associated with Jack's brain,
there are an infinite number of conscious minds.
\end{quote}

Another advocate of many minds is
Lockwood, who first presented his views
soon after Albert and Loewer (Lockwood,
1989, p.~226):
\begin{quote}
According to the relative state view,
there is no collapse (or decomposition)
of the wave function, individual or
multiple.  Rather than say that, on the
relative state view, the observer splits
the universe by carrying out a measurement,
it would be closer to the mark to say that
it is the universe that splits the observer.
\end{quote}
For Lockwood, multiplicity is associated
with higher dimensionality (p.~232):
\begin{quote}
What I am proposing, following Deutsch,
is that we interpret the mathematical
formalism of quantum mechanics
in such a way that the
fact of a physical system's being in a
superposition, with respect to some set
of basis vectors, is to be understood as
the system's having a \emph{dimension} in
addition to those of time and space.
\end{quote}
Lockwood (1996) further developed the
many-minds view and neatly summarized it
(pp.~170--171):
\begin{quote}
A many minds theory, as I understand it,
is a theory which takes completely at face
value the account which unitary quantum
mechanics gives of the physical world and
its evolution over time. In particular, it
allows that, just as in special relativity
there is a fundamental democracy of
Lorentz frames, so in quantum mechanics
there is a fundamental democracy of vector
bases in Hilbert space. In short, it has no
truck with the idea that the laws of physics
prescribe an \emph{objectively} preferred
basis. For a many minds theorist, the
\emph{appearance} of there being a preferred
basis, like the \emph{appearance} of state
vector reduction, is to be regarded as an
illusion. And both illusions can be explained
by appealing to a theory about the way in
which \emph{conscious mentality} relates to
the physical world as unitary quantum
mechanics describes it \mbox{[\ldots]}
Finally, a many minds theory, like a many
worlds theory, supposes that, associated
with a sentient being at any given time,
there is a multiplicity of distinct conscious
points of view. But a many minds theory holds
that it is these conscious points of view or
`minds,' rather than `worlds,' that are to
be conceived as literally dividing or
differentiating over time---or (as is possible
in principle, though unlikely in practice)
\mbox{[\ldots]} fusing or converging.
\end{quote}

Lockwood's ideas can be formalized as follows:
There is a subsystem of the brain which
he calls `Mind' (capitalized).  Within
Mind is an infinite number of minds, and
each of them can have only a single
`maximal experience' at any time.
In the brain's Hilbert space is a
`consciousness basis' of orthonormal
vectors $|\phi_n\rangle$, each one
corresponding to a given maximal
experience type $E_n$.

Let $\rho = \sum_i w_i |\psi_i\rangle
\langle \psi_i|$ be the Mind's density
operator.  One can write $|\psi_i\rangle
= \sum_n c_{in} |\phi_n\rangle$, and
therefore
\begin{equation}
\rho = \sum_{m,n} \left( \sum_i w_i
c_{im} c^*_{in} \right) |\phi_m\rangle
\langle \phi_n| .
\end{equation}
Lockwood postulates that in $\rho$,
$E_n$ occurs with weight
$\sum_i w_i |c_{in}|^2$.
He views that numerical weight as
a segment in a direction orthogonal
to time.  Although he rejects the
transporal identity of individual
minds, he associates segments
corresponding to an instant $t_2$
with segments at $t_1$ through
evolution of the $|\phi_n\rangle$.

According to Lockwood the Mind, in
order to have a single maximal experience,
has to be exactly in a $|\phi_n\rangle$.
If it is described by a slightly different
mixture, it won't have a slightly different
experience, but a small fraction of minds
will rather have vastly different
experiences.  Note that the existence
of an infinite number of minds defined
on a substrate with a finite number of
particles and states (with energy smaller
than some value) seems to introduce a
dualistic component in the theory.

Lockwood's ideas elicited a number of
comments.  Deutsch (1996, p.~224) believes
that the split goes beyond the mind:
\begin{quote}
Lockwood's preference for the term
`many minds' over `parallel universes'
risks giving the impression that it is
\emph{only} minds that are multiple,
and not the rest of reality.  Nothing
could be further from the truth, or
from Lockwood's theory \mbox{[\ldots]}
The distinctive assertion of many-minds
theories is that the universe perceived
by any one mind is not an objectively
separate `layer' of the multiverse.
It is merely the view of the multiverse
from the perspective of that mind.
\end{quote}
On the other hand, Butterfield (1996,
p.~202) pointed out the macroscopic
indefiniteness of Lockwood's view:
\begin{quote}
[I]t is worth emphasizing [Lockwood's
interpretation's] radicalism: it allows
the unobserved macroscopic world to be
very indefinite, even within a branch.
\end{quote}
Earlier, Butterfield (1995, p.~151) 
had already suggested that many minds
``should bite the bullet: the unobserved
rock may well not be definite for any
reasonable familiar quantity.''  Note
that Barrett (1999, p.~210) claims that
``Lockwood's denial of the transcendental
identity of minds \mbox{[\ldots]} makes
his theory empirically incoherent.''

Butterfield (1995, p.~134) points out that
the preferred basis indefiniteness is less
of a problem in many minds than in many
worlds:
\begin{quote}
[M]ost advocates [of many worlds]
have assumed a notion of apparatus,
i.e. a distinguished set of subsystems
of the universe, and a distinguished
quantity on such an apparatus,
e.g.\ position of the apparatus' pointer.
Commentators have often criticized
these assumptions as at best imprecise,
and at worst question-begging
(at least as part of a solution to the
measurement problem) \mbox{[\ldots]}
As we shall see, [many minds] sees itself,
with some justice, as improving on this
imprecise answer, in effect by picking
out as definite those quantities on the brain
that correspond to perception of a definite
pointer-position.
\end{quote}
His view of many minds is rather close to a
modal interpretation (1995, pp.~145--146,
148):
\begin{quote}
[Many minds] is a proposal for how to pick
out the subsystems of the universe,
and the quantities (bases) on them,
that are to define the branches. Roughly
speaking, it proposes that the subsystems
be brains (considered as quantum systems),
and that the quantities be those quantities
whose eigenstates are, or underlie,
conscious mental states.

Then the proposal is: mind or consciousness
picks out a factorization of the universe's
state-space (into the state-space of the brain,
and the state-space of the rest of the universe),
and then also picks out a basis on the first
factor---the basis of those quantum states
that are or underlie conscious mental states.
A branch is defined by an element of the basis,
together with the relative state of the rest of
the universe for that element.
\end{quote}

Bacciagaluppi (202, p.~109) believes that
``the concept of a brain state corresponding
to a definite perception \mbox{[\ldots]}
should not be thought of as providing a
`global' preferred basis for the universe,
but a set of `local' preferred bases, one
for every observer.''

Page (1996, 1997) proposed an approach
related to many minds, which he calls `Sensible
Quantum Mechanics' (SQM).  In the spirit of
Everett, the quantum state never collapses in SQM,
but there is a multiplicity of conscious
perceptions.  Page (1996, p.~585) bases his
approach on three axioms:
\begin{quote}
\textbf{Quantum World Axiom}: The unconscious
`quantum world' $Q$ is completely described by
an appropriate algebra of operators and by a
suitable state $\sigma$ (a positive linear
functional of the operators) giving the expectation
value $\langle O \rangle \equiv \sigma [O]$ of
each operator~$O$.

\textbf{Conscious World Axiom}: The `conscious
world' $M$, the set of all perceptions $p$, has a
fundamental measure $\mu (S)$ for each subset $S$
of $M$.

\textbf{Quantum-Consciousness Connection}: The
measure $\mu (S)$ for each set $S$ of conscious
perceptions is given by the expectation value
of a corresponding `awareness operator' $A(S)$,
a positive-operator-valued (POV) measure
\mbox{[\ldots]}, in the state $\sigma$ of the
quantum world:
\[
\mu (S) = \langle A(S) \rangle
\equiv \sigma [A(S)] .
\]
\end{quote}
In SQM perceptions are basic, and there is no
fundamental way to classify them into individual
persons or minds.  Page leaves it open whether
perceptions, whose measure is determined by the
quantum state, in turn affect the quantum state.
He sees his approach closer to Lockwood's
than to Albert and Loewer's. 

In a series of papers written over a decade,
Donald (1990, 1992, 1995, 1997, 1999) has
developed a technically very sophisticated
theory of many minds.  It is impossible
to summarize it in a few paragraphs, but
some consequences on multiplicity can be
briefly extracted.

Donald presents his views of many minds
as follows (1990, p.~48):
\begin{quote}
[T]he universe exists in some fundamental
state $\omega$. At each time $t$ each observer
$o$ observes the universe, including his own
brain, as being in some quantum state $\sigma_{o,t}$.
Observer $o$ exists in the state $\sigma_{o,t}$,
which is just as `real' as the state $\omega$.
$\sigma_{o,t}$ is determined by the observations
that $o$ has made and, therefore, by the state
of his brain \mbox{[\ldots]} The a priori
probability of an observer existing in state
$\sigma_{o,t}$ is determined by $\omega$.
It is because these a priori probabilities
are predetermined that the laws of physics
and biology appear to hold in the universe
which we observe. According to the
many-worlds theory, there is a huge difference
between the world that we appear to experience
(described by a series of states like $\sigma_{o,t}$)
and the `true' state $\omega$ of the universe.
For example, in this theory, `collapse' is
observer dependent and does not affect $\omega$.
Analysing the appearance of collapse for an
observer is one of the major tasks for the
interpreter of quantum theory. 
\end{quote}

Now one can ask (1990, p.~47),
``What sort of quantum state describes a
brain that is processing definite information
\mbox{[\ldots]}?''  Donald proposes that the
part of the brain relevant to mind can be
modeled by a family of switches.  An observed
phenomenon is then a pattern of switching in a
human brain.  ``[M]ind exists as awareness of brain,
but \mbox{[\ldots]} it has no direct physical effect.''
(1990, p.~53)  However (1992, p.~1133) ``the
(objective) physical substrate of consciousness
\mbox{[\ldots]} is not just the instantaneous
state of a brain but instead involves the
history of that brain.''

In Donald's approach (1992, p.~1149),
``collapse is, one might say, a mistake which
the observer makes about the state of the world
because he is physically incapable of seeing
its true state.''  The number
of worlds is finite, but may be different
from one observer to the next.  Donald's
approach is formulated so as to be
consistent with special relativity and
quantum field theory.  In the end
(1999, p.~19):
\begin{quote}
Nothing is `real' except the switching
structures of individual observers (each
considered separately), the initial condition
$\omega$, the underlying quantum field theory,
and the objective probabilities defined
by the hypothesis. Out of these `elements
of reality,' each separate observer must
construct his experiences and learn to guess
at what his future may bring. This is done by
the observer being aware of his structure as
awareness of an `observed world'. How this
might be possible is \mbox{[\ldots]} a sophisticated
form of the doctrine that one is aware of the
external world entirely through being aware
of the history of one's own brain.
\end{quote}

I should note that there
is sometimes little distinction
between many worlds and many minds, as the
following quotes from Vaidman (1998,
pp.~245, 255, 257--258) illustrate:
\begin{quote}
The `world' is a subjective concept of a
sentient observer.

The basis of the decomposition \mbox{[\ldots]}
of the Universe is determined by the
requirement that individual terms
$|\psi_i\rangle$ correspond to sensible
worlds.  The consciousness of sentient
beings who are attempting to describe
the Universe \emph{defines} this basis.

Every time we encounter a situation in which,
according to the standard approach, collapse
must take place, there splitting in fact takes
place; and the ambiguity connected with the
stage at which collapse occurs corresponds to
the subjective nature of the concept of world.
While this ambiguity represents a very serious
difficulty of the collapse theories, it is not a
serious problem in the MWI.  The collapse
as a physical process should not be vaguely
defined, while the vagueness of the concept
of a conscious being is more of an advantage
than a problem.
\end{quote}
%
%\newpage
\section{Decoherent sectors\label{deco}}
When Everett's theory is construed
as representing a genuine split into
many worlds, the split occurs according
to a specific decomposition of the
total state vector.  That decomposition
is taken to coincide with states where
the apparatus pointer, or more generally
any macroscopic object, is well defined.
Clearly, nothing in the total state vector
singles out such a decomposition.  This
is the preferred-basis problem.

It turns out, however, that this problem
is considerably attenuated if dynamics is
added to the instantaneous state vector.
This is a consequence of decoherence
theory (Schlosshauer, 2004), whose detailed
consideration lies beyond the scope of
this paper.  We will nonetheless say a few
words about it, since approaches to Everett
discussed in this section make essential
use of~it.

Let us for the moment go back to~(\ref{final})
and take $O$ to represent a general environment
instead of a human observer.  Let $\rho$ be
the reduced density operator obtained by tracing
out the environmental degrees of freedom in the
final state.  We then get
\begin{align}
\rho &= |c_1|^2 (|a_1\rangle \langle a_1|)
(|\alpha_1\rangle \langle \alpha_1|)
+ |c_2|^2 (|a_2\rangle \langle a_2|)
(|\alpha_2\rangle \langle \alpha_2|) \notag\\
& \qquad + c_1 c_2^* \langle O_2|O_1 \rangle
(|a_1\rangle \langle a_2|)
(|\alpha_1\rangle \langle \alpha_2|)
+ c_1^* c_2 \langle O_1|O_2 \rangle
(|a_2\rangle \langle a_1|)
(|\alpha_2\rangle \langle \alpha_1|) .
\label{reduced}
\end{align}
It has been shown that in a wide
variety of models, the environment states
$|O_1\rangle$ and $|O_2\rangle$ are nearly
orthogonal if $|\alpha_1\rangle$ and
$|\alpha_2\rangle$ represent macroscopic
states.  The off-diagonal terms in $\rho$
therefore vanish for all practical
purposes.

The fact that the right-hand side
of~(\ref{reduced}) essentially coincides
with the right-hand side of~(\ref{collapse})
has sometimes been taken as an implementation
of the Dirac--von Neumann collapse.  In a
collapse, however, $\rho$ is a so-called proper
mixture, one that represents ignorance
of a true state of affairs that is either
$(|a_1\rangle \langle a_1|) (|\alpha_1\rangle
\langle \alpha_1|)$ or $(|a_2\rangle \langle a_2|)
(|\alpha_2\rangle \langle \alpha_2|)$.
But $\rho$ in~(\ref{reduced})
represents an improper mixture,
which cannot be interpreted as ignorance.

The theory of decoherence turns out
to be an important building block of
an approach to Everett different from many
worlds and many minds.\footnote{The relation
between this more recent approach and
Everett's own views is analyzed in Barrett
(2011, 2014).  Briefly, Everett's
understanding of empirical faithfulness
only requires that the observer and
relative record be found in \emph{some}
decomposition of the state.  There is no
appeal to a preferred basis or to
decoherence considerations.}  The first
step in that direction was taken by
Gell-Mann and Hartle (1990), whose program
``is an attempt at extension,
clarification, and completion of the
Everett interpretation.'' (p.~430)
Gell-Mann and Hartle make use of the
consistent histories formalism.  In the
end however, they take only one world
to be real.\footnote{See also
Seidewitz (2007) and Gell-Mann and
Hartle (2012).}

In a series of papers published
in the 1990s, Saunders (1993, 1995,
1996a, 1996b, 1998) has developed
an approach to Everett based on
decoherence theory.  Saunders begins
by developing an elaborate analogy
between quantum mechanics and special
relativity.\footnote{A similar analogy was
made in Geroch (1984).}  He points out that
most thinkers nowadays view the notion
of `now,' or of `the present,' as
relational rather than absolute.  Indeed
Minkowski's four-dimensional picture of
space-time leaves no room for identifying
something like `now,' or an absolute
past or absolute future.  Time is
relational: one can say (in a given
Lorentz frame for instance) that an instant
$t_2$ is earlier than, simultaneous with
or later than $t_1$, but not that it lies in
some absolute past, present or future.

So it is, according to Saunders, with
actuality in quantum mechanics.  Actuality
is to be viewed as purely relational.
\begin{quote}
\emph{The basic idea of the relational approach
is that this is all that is required at the level
of the fundamental equations.}  What is
`actual,' just as what is `now,' are to be
understood as facts as relations. There is
nothing more to be put in; neither the `flow'
of time, taking us from one `now' to the next,
nor the reduction of state, taking us from one
`actuality' to another.  (1995, p.~243)

Our approach is of Everett type,
qualified as follows: there is a plurality of
`observers' or `classical worlds'
(Everett worlds), but only in the same
Pickwickian sense that there exists
a plurality of spacelike hypersurfaces
(within a fixed spacetime foliation),
a plurality of `nows.'  (1993, p.~1554)
\end{quote}
The formalism of quantum mechanics is to
be based solely on the universal quantum
state and its unitary evolution determined
by some universal Hamiltonian operator.
Suppose that the right-hand side
of~(\ref{final}) represents the universal
state at some time $t$, where $|O_1\rangle$
and $|O_2\rangle$ are (nearly) orthogonal
environment states.  Then
$|a_1\rangle |\alpha_1 \rangle$ is viewed as
actual with respect to $|O_1\rangle$ and
$|a_2\rangle |\alpha_2 \rangle$ is actual
with respect to $|O_2\rangle$.  Neither
the macroscopic pointer state
$|\alpha_1 \rangle$ nor the state
$|\alpha_2 \rangle$ are actual in any
absolute sense, but both are actual in
a relational sense.

Since no basis is singled out in the
universal quantum state, the notion of
actuality is not restricted to
macroscopically well-defined states.
Indeed if~(\ref{final}) is written in
terms of linear combinations $|O'_1\rangle$
and $|O'_2\rangle$ of $|O_1\rangle$
and $|O_1\rangle$, then the state actual
relative to $|O'_1\rangle$, for instance,
will be a linear combination of
$|a_1\rangle |\alpha_1 \rangle$ and
$|a_2\rangle |\alpha_2 \rangle$.

The fact that organisms like humans
perceive macroscopic objects as definite
does not give them any enhanced actuality.
``The goal is to define a sense in which
`the classical' can be understood as an
anthropocentric structure within quantum
mechanics.''  (Saunders, 1993, p.~1560)
Indeed complex structures are taken
to have evolved in the universal
state through something like
Darwinian pressure.  It is a selective
advantage to perceive definite macroscopic
objects, which is why such objects are
actual relative to evolved organisms like
humans.

Saunders (1993) also rejects the
notion of splitting, no more appropriate
than the one of a `now' containing
different times.

Wallace (2002, 2003) follows Saunders
in the analogy between quantum mechanics
and special relativity, as well as on
decoherence and evolution favoring the
emergence of stable structures.  He
identifies real structures with stable
patterns in the universal quantum state
(2003, p.~91):
\begin{quote}
My claim is instead that the emergence
of a classical world from quantum
mechanics is to be understood in terms
of the emergence from the theory of certain
sorts of structures and patterns, and that this
means that we have no need (as well as no
hope!) of the precision which Kent and others
here demand.
\end{quote}
Going back to~(\ref{final}) once again,
there is one apparatus pattern
$|\alpha_0 \rangle$ in the universal
quantum state at the beginning of the
experiment.  Therefore there is one
apparatus.  At the end of the experiment,
the universal quantum state contains
two distinct apparatus patterns
$|\alpha_1 \rangle$ and $|\alpha_2 \rangle$.
Therefore there are now two apparatus.
According to Wallace (2003, p.~92),
this doesn't cause problems because
\begin{quote}
If A and B are to be `live cat' and `dead cat'
then P and Q will be described by statements
about the state vector which (expressed in a
position basis) will concern the wave-function's
amplitude in vastly separated regions $R_P$ and
$R_Q$ of configuration space, and there will
be no contradiction between these statements.
\end{quote}
Just like Saunders, Wallace allows that
worlds are imprecisely defined:
\begin{quote}
The problem given rise to by this abstraction
is that there exist many choices of consistent
history space, but if we follow Everett and
keep the state as fundamental there is no
problem. Just as our choice of
world-decomposition (i.e.\ fine-grained basis)
is made for ease of description rather than
more fundamental reasons, so our choice
of history space is just made so as to give a
convenient description of the quantum universe.
(2002, pp.~648--649)

Another question which at first sight should
have a precise answer: if there was one cat
before the measurement and two after it, when
\emph{exactly} did the duplication of cats occur?
\mbox{[\ldots]} Put another way, the cat
description is only useful when answering
questions on timescales far longer than [the
decoherence timescale] $\tau_D$, so \emph{whether
or not} quantum splitting is occurring,
it just doesn't make sense to ask questions about
cats that depend on such short timescales.
(2003, pp.~97--98)
\end{quote}

We should point out that the present
approach views the evolution of
Schr\"{o}\-dinger's cats quite differently
than the many-worlds approach.  Let
$|\mbox{Live}\rangle$ and $|\mbox{Dead}\rangle$ 
represent states of the live and dead cat,
and let $|\mbox{ND}\rangle$ and
$|\mbox{D}\rangle$ represent states of
the nondecayed and decayed nucleus.
The compound system starts
at $t=0$ in the state
$|\mbox{ND}\rangle |\mbox{Live}\rangle$.
At time $t$, the state vector has become
\begin{equation}
|\psi (t)\rangle = e^{-\alpha t}
|\mbox{ND}\rangle |\mbox{Live}\rangle
+ \sqrt{1 - e^{- 2 \alpha t}} |\mbox{D}\rangle
|\mbox{Dead}\rangle ,
\label{cat}
\end{equation}
where $\alpha^{-1} \ln 2$ is the nucleus'
half-life.  In the present approach,
(\ref{cat}) means that the measure of
worlds where the cat is dead continuously
increases.  In the many-worlds approach,
there must be a continuous split of the
worlds described by the first term on
the right-hand side of~(\ref{cat}), so
as to constantly increase the number
of worlds described by the second term.

It can be argued that nonlocality
is less of a problem in decoherence
than in many-worlds approaches.
Referring to EPR-type experiments,
Hewitt-Horsman (2009, p.~888) concludes
that
\begin{quote}
the second particle is still
correlated with the first in neo-Everett,
and at first glance it would seem that some
sort of non-local signal \emph{is} needed,
to tell the worlds of the first particle
how they join up with the worlds of the
second particle.  This is indeed necessary
in those many world theories that have
spatially extended worlds that split
instantaneously.
\end{quote}

Butterfield (2001, p.~133) has pointed out the
rather complex ontology of Saunders'
and Wallace's approach:
\begin{quote}
Saunders and Wallace conclude from these
difficulties that we should liberally
accept resolutions of the universal
quantum state $\Psi$ into an
arbitrary basis---or at least an arbitrary
basis that is a fine-graining of `the'
decoherence basis. In
Wallace's terminology of `worlds,'
they consider continuously many bases
(even if they restrict themselves to
bases that fine-grain `the' decoherence basis),
and so commit themselves to continuously
many worlds \mbox{[\ldots]}
This is certainly a dizzying ontology. After all,
each of these continuously many worlds is
`inhabited.'  Each world is not just a
component of a state (where states represent
reality) but also has a `system' (albeit not
an ordinary object!) actually in it.
They would still have continuously many worlds
even if for each basis they said---as they
do not---that only one world is `inhabited.'
\end{quote}

Kent (2010, p.~311) argues that the
decoherence approach is close to the
many-minds approach:
\begin{quote}
[I]t seems to me \mbox{[\ldots]} that,
at various points in their arguments,
Saunders, Wallace, Greaves--Myrvold
and Papineau tacitly---and, since they
reject the many-minds interpretation,
illegitimately---appeal to many-minds
intuitions.  Indeed, at least in the
first three cases, it seems to me that
if one fleshed their ideas out into a
fully coherent and complete interpretation,
one would necessarily arrive either at the
many-minds interpretation or something
even worse.
\end{quote}
But neither Saunders nor Wallace want
to interpret multiplicity in terms
of many worlds or many minds.  How then
are they going to interpret it?

Let us focus on Wallace's patterns
and, for vividness, formulate the question
in terms of Schr\"{o}dinger's cat.  Wallace
claims that there is no contradiction in
the simultaneous existence of the live cat
and the dead cat.  The reason is that they
both correspond to well-defined patterns
in the universal quantum state and that
the two patterns are associated with vastly
different regions in configuration space.  
This is also what happens in classical
theory, where two different cats will
always occupy vastly different regions in
configuration space.

But there is a crucial difference between
classical patterns and patterns in the
universal state vector.  In classical
theory, two different cats will not only
occupy different regions in configuration
space, but they will also occupy disjoint
regions in three-dimensional space.  That
is, their projections from configuration
space to three-space will not overlap.
This is not so with Wallace's patterns.
Projected in three-space, the live cat
may not only step on the dead cat's tail,
he may literally get into it.  How can
one understand such multiplicity?

First note that, according to Wallace
(2012, p.~40), ``cats, or tables, or any
other such entities exist \mbox{[\ldots]}
as structures within the underlying
microphysics.''  In principle, a cat could
be described by applying molecular
dynamics to its constituent particles,
but this is utterly impracticable as
well as not particularly illuminating.
Higher-level structures such as
organelles, cells, tissues and organs
can be useful to study different
characteristics of a cat.  At the
highest level, patterns elucidated by
zoology and evolutionary biology will
help us understand, for instance, the
hunting behavior of cats.

Whatever the relevance of studying each
structure or pattern at its own level,
Wallace allows that cats are made of organs,
organs of tissues, tissues of cells,
etc., all the way down to
molecules and atoms.  Let us focus for a
moment on the level of DNA molecules.
These are rather well understood from an
atomic point of view, but mostly behave
quasi-classically.  Specifically, they
are stable over long periods of time,
witness the fact that even prehistoric
DNA can often be sequenced.  Most
importantly, DNA molecules of
Schr\"{o}dinger's cat are stable throughout
the fate of the unfortunate animal.

So if the live cat and the dead cat are
both real, each has its own DNA molecules.
Therefore as the experiment proceeds, there
come to be twice as many DNA molecules as
before.  The same remark also applies,
it seems, to the atomic constituents of
DNA, but let us stick to DNA itself.  The
question is, where in three-dimensional
space are all these molecules?

One possible answer is to assume that
the live cat and the dead cat, and their
associated DNA, don't project from
configuration space to the same three-space.
This is suggested, for instance, by
Bacciagaluppi (2002, p.~118):
``[I]t should be possible to show that
the total set of events will not fit into
one simple space-time, but into a
\textit{branching} space-time. The branching
space-time ought to be reconstructed from
the causal structure of the decoherence
events.''  This means that there is an
added parameter, or another dimension,
introduced to distinguish different
three-spaces from each other.  It is hard
to see, however, what difference there is
between this scheme and many worlds.

But this is not Wallace's answer.
According to him (2012, p.~311),
there is only one three-space into
which both patterns project:
``Other branches, then, are located in
precisely the same space and time as our
own world; it is just that they are
dynamically incapable of (significantly)
affecting our world, and vice versa.''
DNA molecules of the live and dead cats
are, so to speak, ghostlike to each other.
More macroscopically, the live cat's
paw indeed gets into the dead cat's tail,
but he is entirely unaware of it.  Or,
as Allori \emph{et al.} put it:
\begin{quote}
Note that, by the linearity of the Schr\"{o}dinger
evolution, the live cat and the dead cat,
that is $m_1$ and $m_2$, do not interact with
each other, as they correspond to $\psi_1$ and
$\psi_2$, which would in the usual quantum
theory be regarded as alternative states of
the cat. The two cats are, so to speak,
reciprocally transparent. (2011, p.~7)

Metaphorically speaking, the universe
according to Sf or Sm resembles the situation
of a TV set that is not correctly tuned, so
that one always sees a mixture of two channels.
(2008, p.~379)
\end{quote}
It is not clear how microphysics can be
adapted for particles in the same three-space
to only selectively interact with each other.

Do we definitely have to choose between
one three-space and many three-spaces?
Maybe not.  Wilson (2011, p.~375) suggests
that
\begin{quote}
the `spacetime' of the quantum mechanics
and quantum field theory formalism,
in terms of which branches are defined,
is not the same as the `spacetimes' of
macroscopic worlds. The former `spacetime'
is a single entity common to multiple
branches, while each of the latter
`spacetimes' is tied to a particular
macroscopic course of events.
\end{quote}
The idea may be worth investigating,
but its full implementation may not
be easily carried out.

I should note that not everyone agrees
with Wallace on the reducibility, in
principle, of macroscopic structure to
the underlying microphysics.  In
developing what they call `ontic
structural realism,' Ladyman and Ross
(2007) argue that structure is more
real than objects, and that patterns
at one level are not made out of more
fundamental objects.  Can Everett's
approach be formulated within this
type of metaphysics?  Maybe, but Ladyman
and Ross are not themselves advocates
of Everett's approach.

If one follows Saunders and accepts
that actuality is relational, perhaps
then space itself is relational.
Again, this would require new physics,
since both in nonrelativistic quantum
mechanics and in quantum field theory,
the space-time arena is presupposed
antecedently.  Three-dimensional space
cannot be taken as emerging solely from
the wave function, since all separable
Hilbert spaces are isomorphic.
%
%\newpage
\section{Discussion\label{disc}}
\begin{quote}
After 50 years, there is no well-defined,
generally agreed set of assumptions and
postulates that together constitute
`the Everett interpretation of quantum
theory.' (Kent, 2010, p.~310)
\end{quote}
This assessment, I believe, has been
substantiated in this paper with respect
to multiplicity:  Are worlds physically
splitting, or does the split happen only in
the mind?  Do worlds split locally or
globally, instantaneously or on some other
space-time hypersurface?  Are worlds
generated upon any entanglement-producing
interaction, or only when macroscopic
objects or apparatus are involved?  Do
worlds split or bifurcate?  Do they occupy
the same physical space-time or do they
involve extra dimensions?  These and others
are all questions on which different
investigators disagree, or even questions
that have received different answers at
different times from the same investigator.
This, I should stress, is not meant as a
charge of inconsistency, since opinions
and points of view naturally evolve in
the course of research.  But it illustrates
how long the road still is to a full
clarification of Everett's interpretation.

I am much attracted by the so-called
semantic view of theories which, as
far as quantum mechanics is concerned,
construes interpretation as answering the
question ``How can the world be for
quantum mechanics to be true?''
From such a point of view, dealing with
different consistent interpretations of a
theory provides clarification and
understanding rather than confusion
(Marchildon 2004, 2009).  There is no doubt
that Everett's relative states approach
constitutes a major contribution to
the debate on the interpretation of
quantum mechanics.  Every interpretation
of quantum mechanics has its shortcomings,
since none has yet rallied general
support.  But in closing I would like
to stress a few problems that are perhaps
more specific to Everett's theory.

The first problem is what has been documented
here, that Everett's theory is not well defined.
Such a charge can also be levelled at the
Copenhagen interpretation or at the
Dirac--von Neumann collapse postulate.
This is in sharp contrast with the
de Broglie--Bohm approach which,
at least in the nonrelativistic case,
is precisely defined.  I also believe
that the more recent approach of
Quantum Bayesianism (Fuchs \emph{et al.},
2014) is well defined, although it has
problems of its own (Marchildon, 2015).
In all fairness, the de Broglie--Bohm
approach also has yet unsolved problems
in the relativistic or field-theoretic case.

What I see as the second problem
with Everett's approach may well be
viewed as a virtue by its advocates:
it is the fact that the approach
stands or falls with the exact
validity of quantum mechanics.
Add the smallest nonlinear term to
the Schr\"{o}dinger equation, and
Everett's many worlds disappear just
like rings of smoke.  The provisional
status of fundamental theories suggests
that wide-ranging ontological claims
should not depend on such strong
assumptions.  By contrast, the
de Broglie-Bohm approach is highly
adaptable to changes in the formalism
of quantum mechanics (Valentini, 2010).

Finally, I cannot help comparing
Everett's extraordinary ontology with
other such wildly counter-intuitive
instances in the history of thought.
I have in mind, for instance,
Parmenides' rejection of motion to 
match a theory of being, or Berkeley's
rejection of matter to avoid the
mind-body problem.  In experimental
science, Carl Sagan's motto that
``Extraordinary claims require
extraordinary evidence'' is a good
methodological attitude.  There is a lesson
to be drawn from this even if the interpretation
of quantum mechanics goes much beyond
experiment.  It seems to me
that much skepticism about Everett's approach
and its many implementations stems from the
fact that in spite of valiant attempts
(Deutsch, 1997), the strength of arguments in
its favor does not match the scope of
its claim.
\section*{Acknowledgments}
This work is supported by the Natural Sciences
and Engineering Research Council of Canada
(RGPIN/5022-2011).
%
%\newpage
\section*{References}
Albert, D. (1986).
How to take a photograph of another
Everett world.
In D. M. Greenberger (Ed.),
\textit{New techniques and ideas in quantum
measurement theory} (pp.~498--502).
Annals of the New York Academy of
Sciences, Vol.~480.\\
\verb+doi:10.1111/j.1749-6632.1986.tb12452.x+.

\bibspace\noindent
Albert, D. (1992).
\textit{Quantum mechanics and experience}.
Cambridge, MA: Harvard University Press.

\bibspace\noindent
Albert, D., \& Barrett, J. A. (1995).
On what it takes to be a world.
\textit{Topoi, 14}, 35--37.
\verb+doi:10.1007/BF00763476+.

\bibspace\noindent
Albert, D., \& Loewer, B. (1988).
Interpreting the many worlds interpretation.
\textit{Synthese, 77}, 195--213.
\verb+doi:10.1007/BF00869434+.

\bibspace\noindent
Allori, V., Goldstein, S., Tumulka, R.,
\& Zanghi, N. (2008).
On the common structure of Bohmian mechanics
and the Ghirardi--Rimini--Weber theory.
\textit{British Journal for the Philosophy
of Science, 59}, 353--389.\\
\verb+doi:10.1093/bjps/axn012+.

\bibspace\noindent
Allori, V., Goldstein, S., Tumulka, R.,
\& Zanghi, N. (2011).
Many worlds and Schr\"{o}dinger's first quantum theory.
\textit{British Journal for the Philosophy
of Science, 62}, 1--27.
\verb+doi:10.1093/bjps/axp053+.

\bibspace\noindent
Bacciagaluppi, G. (2002).
Remarks on space-time and locality
in Everett's interpretation.
In T. Placek, \& J. Butterfield (Eds.),
\textit{Non-locality and modality}
(pp.~105--122).  Dordrecht: Kluwer.\\
\verb+doi:10.1007/978-94-010-0385-8_7+.

\bibspace\noindent
Baker, D. J. (2007).
Measurement outcomes and probability
in Everettian quantum mechanics.
\textit{Studies in History and Philosophy
of Modern Physics, 38}, 153--169.
\verb+doi:10.1016/j.shpsb.2006.05.003+.

\bibspace\noindent
Ballentine, L. E. (1973).
Can the statistical postulate of quantum theory
be derived?---A critique of the many-universes
interpretation.
\textit{Foundations of Physics, 3}, 229--240.
\verb+doi:10.1007/BF00708440+.

\bibspace\noindent
Ballentine, L. E., Pearle, P., Walker,
E. H., Sachs, M., Koga, T., Gerver,
J., \& DeWitt, B. S. (1971).
Quantum-mechanics debate.
\textit{Physics Today, 24}(4), 36--44.
\verb+doi:10.1063/1.3022676+.

\bibspace\noindent
Barrett, J. A. (1995).
The single-mind and many-minds versions
of quantum mechanics.
\textit{Erkenntnis, 42}, 89--105.
\verb+doi:10.1007/BF01666813+.

\bibspace\noindent
Barrett, J. A. (1999).
\textit{The quantum mechanics of minds
and worlds}.
Oxford: Oxford University Press.

\bibspace\noindent
Barrett, J. A. (2011).
Everett's pure wave mechanics
and the notion of worlds.
\textit{European Journal for Philosophy
of Science, 1}, 277--302.\\
\verb+doi:10.1007/s13194-011-0023-9+.

\bibspace\noindent
Barrett, J. A. (2014).
Everett's relative-state formulation
of quantum mechanics.
In \textit{Stanford Encyclopedia of Philosophy}.\\
http://plato.stanford.edu/entries/qm-everett/

\bibspace\noindent
Barrett, J. A., \& Byrne, P. (Eds.) (2012).
\textit{The Everett interpretation of
quantum mechanics.  Collected works
1955--1980 with commentary}.
Princeton: Princeton University Press.

\bibspace\noindent
Bell, J. S. (1976).
The measurement theory of Everett and
de Broglie's pilot wave.
Reprinted in Bell (1987), pp.~93--99.\\
\verb+doi:10.1007/978-94-010-1440-3_2+.

\bibspace\noindent
Bell, J. S. (1981).
Quantum mechanics for cosmologists.
Reprinted in Bell (1987), pp.~117--138.

\bibspace\noindent
Bell, J. S. (1986).
Six possible worlds of quantum mechanics.
Reprinted in Bell (1987), pp.~181--195.

\bibspace\noindent
Bell, J. S. (1987).
\textit{Speakable and unspeakable in quantum mechanics}.
Cambridge: Cambridge University Press.

\bibspace\noindent
Bostr\"{o}m, K. J. (2012).
Combining Bohm and Everett: Axiomatics
for a standalone quantum mechanics.
Preprint arXiv:1208.5632v4.

\bibspace\noindent
Bostr\"{o}m, K. J. (2015).
Quantum mechanics as a deterministic
theory of a continuum of worlds.
\textit{Quantum Studies: Mathematics and
Foundations, 2}, 315--347.
\verb+doi:10.1007/s40509-015-0046-6+.

\bibspace\noindent
Butterfield, J. (1995).
Worlds, minds and quanta.
\textit{Proceedings of the Aristotelian Society,
Supplementary Volumes, 69}, 113--158.

\bibspace\noindent
Butterfield, J. (1996).
Whither the minds?
\textit{British Journal for the Philosophy
of Science, 47}, 200--221.

\bibspace\noindent
Butterfield, J. (2001).
Some worlds of quantum theory.
In R. J. Russell, P. Clayton,
K. Wegter-McNelly, \& J. Polkinghorne (Eds.),
\textit{Quantum mechanics. Scientific
perspectives on divine action} (pp.~111--140).
Vatican Observatory Publications.

\bibspace\noindent
Byrne, P. (2010).
\textit{The many worlds of Hugh Everett III}.
Oxford: Oxford University Press.

\bibspace\noindent
Cooper, L. N., \& Van Vechten, D. (1969).
On the interpretation of measurement
within the quantum theory.
\textit{American Journal of Physics, 37}, 1212--1220.
\verb+doi:10.1119/1.1975279+.

\bibspace\noindent
Deutsch, D. (1985).
Quantum theory as a universal physical theory.
\textit{International Journal of Theoretical
Physics, 24}, 1--41.\\
\verb+doi:10.1007/BF00670071+.

\bibspace\noindent
Deutsch, D. (1996).
Comment on Lockwood.
\textit{British Journal for the Philosophy of
Science, 47}, 222--228.

\bibspace\noindent
Deutsch, D. (1997).
\textit{The fabric of reality}.
London: Penguin.

\bibspace\noindent
DeWitt, B. S. (1970).
Quantum mechanics and reality.
\textit{Physics Today, 23}(9), 30--35.
\verb+doi:10.1063/1.3022331+.

\bibspace\noindent
DeWitt, B. S., \& Graham, N. (Eds.), (1973).
\textit{The many-worlds interpretation
of quantum mechanics}.
Princeton: Princeton University Press.

\bibspace\noindent
Donald, M. J. (1990).
Quantum theory and the brain.
\textit{Proceedings of the Royal Society of London.
Series A, Mathematical and Physical
Sciences, 427}, 43--93.
\verb+doi:10.1098/rspa.1990.0004+.

\bibspace\noindent
Donald, M. J. (1992).
\textit{A priori} probability and
localized observers.
\textit{Foundations of Physics, 22}, 1111--1172.
\verb+doi:10.1007/BF00732696+.

\bibspace\noindent
Donald, M. J. (1995).
A mathematical characterization of the physical
structure of observers.
\textit{Foundations of Physics, 25}, 529--571.\\
\verb+doi:10.1007/BF02059006+.

\bibspace\noindent
Donald, M. J. (1997).
On many-minds interpretations of
quantum theory.
Preprint arXiv:quant-ph/9703008v2.

\bibspace\noindent
Donald, M. J. (1999).
Progress in a many-minds interpretation
of quantum theory.
Preprint arXiv:quant-ph/9904001v1.

\bibspace\noindent
Everett III, H. (1957a).
`Relative state' formulation of quantum mechanics.
\textit{Reviews of Modern Physics, 29}, 454--462.\\
\verb+doi:10.1103/RevModPhys.29.454+.

\bibspace\noindent
Everett III, H. (1957b).
The theory of the universal wave function.
Reprinted in DeWitt \& Graham (1973), pp.~3--140.

\bibspace\noindent
Fuchs, C. A., Mermin, N. D., \& Schack, R. (2014).
An introduction to QBism with an application
to the locality of quantum mechanics.
\textit{American Journal of Physics,
82}, 749--754.
\verb+doi:10.1119/1.4874855+.

\bibspace\noindent
Gauthier, Y. (1983).
Quantum mechanics and the local observer.
\textit{International Journal of Theoretical
Physics, 22}, 1141--1152.\\
\verb+doi:10.1007/BF02080320+.

\bibspace\noindent
Gell-Mann, M., \& Hartle, J. B. (1990).
Quantum mechanics in the light of quantum
cosmology.
In W. H. Zurek (Ed.),
\textit{Complexity, entropy, and the physics
of information} (pp.~425--458).
Reading, MA: Addison-Wesley.

\bibspace\noindent
Gell-Mann, M., \& Hartle, J. B. (2012).
Decoherent histories quantum mechanics with
one `real' fine-grained history.
\textit{Physical Review A, 85}, 062120.\\
\verb+doi:10.1103/PhysRevA.85.062120+.

\bibspace\noindent
Geroch, R. (1984).
The Everett interpretation.
\textit{No\^{u}s, 18}, 617--633.\\
\verb+doi:10.2307/2214880+.

\bibspace\noindent
Graham, N. (1973).
The measurement of relative frequency.
In DeWitt \& Graham (1973), pp.~229--253.

\bibspace\noindent
Healey, R. A. (1984).
How many worlds?
\textit{No\^{u}s, 18}, 591--616.\\
\verb+doi:10.2307/2214879+.

\bibspace\noindent
Hewitt-Horsman, C. (2009).
An introduction to many worlds in
quantum computation.
\textit{Foundations of Physics, 39}, 869--902.\\
\verb+doi:10.1007/s10701-009-9300-2+.

\bibspace\noindent
Jekni\'{c}-Dugi\'{c}, J., Dugi\'{c}, M.,
\& Francom, A. (2014).
Quantum structures of a model-universe:
An inconsistency with Everett interpretation
of quantum mechanics.
\textit{International Journal of Theoretical
Physics, 53}, 169--180.\\
\verb+doi:10.1007/s10773-013-1794-x+.

\bibspace\noindent
Kent, A. (1990).
Against many-worlds interpretations.
\textit{International Journal of Modern
Physics A, 5}, 1745--1762;
See also arXiv:gr-qc/9703089.\\
\verb+doi:10.1142/S0217751X90000805+.

\bibspace\noindent
Kent, A. (2010).
One world versus many: The inadequacy of
Everettian accounts of evolution, probability,
and scientific confirmation.
In Saunders \emph{et al.} (2010), pp.~307--354.

\bibspace\noindent
Kent, A. (2012).
Real world interpretations of quantum theory.
\textit{Foundations of Physics, 42}, 421--435.
\verb+doi:10.1007/s10701-011-9610-z+.

\bibspace\noindent
Ladyman, J., \& Ross, D. (2007).
\textit{Every thing must go:
Metaphysics naturalized}.
Oxford: Oxford University Press.

\bibspace\noindent
Lockwood, M. (1989).
\textit{Mind, brain and the quantum:
The compound `I'}.
Oxford: Blackwell.

\bibspace\noindent
Lockwood, M. (1996).
`Many minds' interpretations of quantum mechanics.
\textit{British Journal for the Philosophy
of Science, 47}, 159--188.

\bibspace\noindent
Marchildon, L. (2004).
Why should we interpret quantum mechanics?
\textit{Foundations of Physics, 34},
1453--1466.\\
\verb+doi:10.1023/B:FOOP.0000044100.95918.b2+.

\bibspace\noindent
Marchildon, L. (2009).
Quantum mechanics needs interpretation.
\textit{International Journal on Advances in
Systems and Measurements, 2},
131--141.

\bibspace\noindent
Marchildon, L. (2011).
Can Everett be interpreted without extravaganza?
\textit{Foundations of Physics, 41}, 357--362.
\verb+doi:10.1007/s10701-010-9415-5+.

\bibspace\noindent
Marchildon, L. (2015).
Why I am not a QBist.
\textit{Foundations of Physics, 45}, 754--761.
\verb+doi:10.1007/s10701-015-9875-8+.

\bibspace\noindent
Osnaghi, S., Freitas, F., \& Freire Jr., O. (2009).
The origin of the Everettian heresy.
\textit{Studies in History and Philosophy
of Modern Physics, 40}, 97--123.
\verb+doi:10.1016/j.shpsb.2008.10.002+.

\bibspace\noindent
Page, D. N. (1996).
Sensible quantum mechanics: Are probabilities
only in the mind?
\textit{International Journal of Modern
Physics~D, 5}, 583--596.\\
\verb+doi:10.1142/S0218271896000370+.

\bibspace\noindent
Page, D. N. (1997).
Sensible quantum mechanics: Are only perceptions
probabilistic?
Preprint arXiv: quant-ph/9506010v2.

\bibspace\noindent
Saunders, S. (1993).
Decoherence, relative states, and
evolutionary adaptation.
\textit{Foundations of Physics, 23}, 1553--1585.
\verb+doi:10.1007/BF00732365+.

\bibspace\noindent
Saunders, S. (1995).
Time, quantum mechanics, and decoherence.
\textit{Synthese, 102}, 235--266.
\verb+doi:10.1007/BF01089802+.

\bibspace\noindent
Saunders, S. (1996a).
Time, quantum mechanics, and tense.
\textit{Synthese, 107}, 19--53.
\verb+doi:10.1007/BF00413901+.

\bibspace\noindent
Saunders, S. (1996b).
Comment on Lockwood.
\textit{British Journal for the Philosophy
of Science, 47}, 241--248.

\bibspace\noindent
Saunders, S. (1998).
Time, quantum mechanics, and probability.
\textit{Synthese, 114}, 373--404.
\verb+doi:10.1023/A:1005079904008+.

\bibspace\noindent
Saunders, S. (2010).
Chance in the Everett interpretation.
In Saunders \emph{et al.} (2010), pp.~181--205.

\bibspace\noindent
Saunders, S., Barrett, J., Kent, A.,
\& Wallace, D. (Eds.), (2010).
\textit{Many worlds? Everett,
quantum theory, and reality}.
Oxford: Oxford University Press.

\bibspace\noindent
Schlosshauer, M. (2004).
Decoherence, the measurement problem,
and interpretations of quantum mechanics.
\textit{Reviews of Modern Physics, 76},
1267--1305.
\verb+doi:10.1103/RevModPhys.76.1267+.

\bibspace\noindent
Schwindt, J. M. (2012).
Nothing happens in the universe of the
Everett interpretation.
Preprint arXiv:1210.8447v1.

\bibspace\noindent
Seidewitz, E. (2007).
The universe as an eigenstate:
Spacetime paths and decoherence.
\textit{Foundations of Physics, 37}, 572--596.\\
\verb+doi:10.1007/s10701-007-9123-y+.

\bibspace\noindent
Smolin, L. (1984).
On quantum gravity and the many-worlds
interpretation of quantum mechanics.
In S. M. Christensen (Ed.),
\textit{Quantum theory of gravity}
(pp.~431--454).  Bristol: Adam Hilger.

\bibspace\noindent
Squires, E. J. (1987).
Many views of one world---An
interpretation of quantum theory.
\textit{European Journal of Physics, 8},
171--173

\bibspace\noindent
Squires, E. J. (1988).
The unique world of the Everett version
of quantum theory.
\textit{Foundations of Physics Letters, 1},
13--20.\\
\verb+doi:10.1007/BF00661313+.

\bibspace\noindent
Squires, E. J. (1991).
One mind or many---A note on the
Everett interpretation of quantum theory.
\textit{Synthese, 89}, 283--286.\\
\verb+doi:10.1007/BF00413909+.

\bibspace\noindent
Tappenden, P. (2000).
Identity and probability in Everett's
multiverse.
\textit{British Journal for the Philosophy
of Science, 51}, 99--114.\\
\verb+doi:10.1093/bjps/51.1.99+.

\bibspace\noindent
Tipler, F. J. (1986).
The many-worlds interpretation of quantum
mechanics in quantum cosmology.
In R. Penrose, \& C. J. Isham (Eds.),
\textit{Quantum concepts in space and time}
(pp.~204--214).  Oxford: Clarendon Press.

\bibspace\noindent
Tipler, F. J. (2006).
What about quantum theory?
Bayes and the Born interpretation.
Preprint arXiv:quant-ph/0611245v1.

\bibspace\noindent
Vaidman, L. (1998).
On schizophrenic experiences of the neutron
or why we should believe in the many-worlds
interpretation of quantum mechanics.
\textit{International Studies in the
Philosophy of Science, 12}, 245--261.\\
\verb+doi:10.1080/02698599808573600+.

\bibspace\noindent
Vaidman, L. (2014).
Many-worlds interpretation of quantum mechanics.
In \textit{Stanford Encyclopedia of Philosophy}.\\
http://plato.stanford.edu/entries/qm-manyworlds/.

\bibspace\noindent
Valentini, A. (2010).
De Broglie--Bohm pilot-wave theory:
Many worlds in denial?
In Saunders \emph{et al.} (2010), pp.~476--509.

\bibspace\noindent
Vermaas, P. (1999).
\textit{A philosopher's understanding of quantum
mechanics.  Possibilities and impossibilities
of a modal interpretation}.
Cambridge: Cambridge University Press.

\bibspace\noindent
Von Neumann, J. (1955).
\textit{Mathematical foundations of quantum mechanics}.
Princeton: Princeton University Press.

\bibspace\noindent
Wallace, D. (2002).
Worlds in the Everett interpretation.
\textit{Studies in History and Philosophy of
Modern Physics, 33}, 637--661.\\
\verb+doi:10.1016/S1355-2198(02)00032-1+.

\bibspace\noindent
Wallace, D. (2003).
Everett and structure.
\textit{Studies in History and Philosophy of
Modern Physics, 34}, 87--105.
\verb+doi:10.1016/S1355-2198(02)00085-0+.

\bibspace\noindent
Wallace, D. (2012).
\textit{The emergent multiverse. Quantum
theory according to the Everett interpretation}.
Oxford: Oxford University Press.

\bibspace\noindent
Whitaker, M. A. B. (1985).
The relative states and many-worlds
interpretations of quantum mechanics
and the EPR problem.
\textit{Journal of Physics A: Mathematical
and General, 18}, 253--264.\\
\verb+doi:10.1088/0305-4470/18/2/015+.

\bibspace\noindent
Wilson, A. (2011).
Macroscopic ontology in Everettian
quantum mechanics.
\textit{Philosophical Quarterly, 61}, 363--382.\\
\verb+doi:10.1111/j.1467-9213.2010.675.x+.

\bibspace\noindent
Zeh, H. D. (1970).
On the interpretation of measurement in quantum theory.
\textit{Foundations of Physics, 1}, 69--76.
\verb+doi:10.1007/BF00708656+.

\bibspace\noindent
Zeh, H. D. (1981).
The problem of conscious observation in
quantum mechanical description.
\textit{Epistemological Letters}
(Ferdinand--Gonseth Association), 63.0;
reprinted with additions in
\textit{Foundations of Physics Letters, 13},
221--233 (2000).
\verb+doi:10.1023/A:1007895803485+.

\bibspace\noindent
Zeh, H. D. (2014).
John Bell's varying interpretations
of quantum mechanics.
Preprint arXiv:1402.5498v7.
\end{document}